\documentclass[12pt]{article}
\usepackage{enumitem}
\usepackage[english]{babel}
\usepackage[letterpaper,top=2cm,bottom=2cm,left=3cm,right=3cm,marginparwidth=1.75cm]{geometry}
\usepackage{amsmath}
\usepackage{graphicx}
\usepackage{amsbsy}
\usepackage{amsmath, amsthm, amssymb}
\usepackage{makeidx}
\usepackage{xcolor}
\usepackage{pifont}
\usepackage{longtable}
\usepackage{babel}
\usepackage{subcaption}
\usepackage{natbib}
\usepackage{booktabs}
\usepackage{float}
\usepackage{tikz}
\usepackage{algorithm}
\usepackage[colorlinks=true, allcolors=blue]{hyperref}

\DeclareMathOperator*{\argmax}{argmax}
\usepackage{multirow}
\usepackage{caption}
\usepackage{authblk}

\title{Predicting the 2026 FIFA World Cup with\\
  Sufficient Dimension Reduction of Elo Rating Histories}

\author[]{Mina Rezaei\thanks{Corresponding author: \href{mailto:mina.rezaei@siu.edu}{mina.rezaei@siu.edu}}~}
\author[]{S. Yaser Samadi }
\affil{School of Mathematical and Statistical Sciences, Southern Illinois University Carbondale, IL, USA}
\date{}

\begin{document}
\maketitle

\section*{Abstract}
We study probabilistic forecasting of the 2026 FIFA World Cup, the first edition with 48 teams and an added Round of 32. The main idea is to describe team strength not only by the current Elo rating, but by a short history of recent Elo differences. We then reduce this history to a few informative directions using categorical sufficient dimension reduction (SDR). The reduced scores are used in a Poisson double-regression model for home and away goals, which gives full outcome probabilities.
We compare eleven models, including logistic regression, standard Poisson regression, ARIMA, and neural-network forecasts of the Elo series, gradient boosting, an ensemble model, and four categorical SDR variants based on sliced inverse regression (SIR) and sliced average variance estimation (SAVE). The models are evaluated out of sample on the 2018 and 2022 World Cups using the ranked probability score (RPS). The results show that SDR-based poisson models improve the traditional approaches, suggesting that recent Elo history contains useful predictive information that is not captured by the current Elo difference alone.

\medskip
\noindent\textbf{Keywords:} football forecasting; ranked probability
score; sufficient dimension reduction; Elo ratings; Poisson
regression; FIFA World Cup 2026.

\section{Introduction}
\label{sec:intro}
The FIFA World Cup is the most watched sporting event in the world.
The 2026 edition, hosted jointly by the United States, Canada, and
Mexico (11~June--19~July 2026), introduces a historic change: the
tournament expands from 32 to \textbf{48 teams} in 12 groups of four,
which creates a new Round of 32 and 104 matches in total. The new format has more teams and one extra knockout round, so the tournament is harder to predict. Because of this, it is better to give probabilities for different outcomes instead of choosing only one winner.

Probabilistic prediction of football outcomes has a long history.
One tradition models the goals scored by each team as Poisson random
variables \citep{maher1982,dixon1997,karlis2003}; a second summarises
team strength with an evolving rating, of which the Elo system
\citep{elo1978,hvattum2010} is the most widely used; and a third
applies machine-learning methods such as gradient boosting
\citep{groll2019,hubacek2019}. These approaches are reviewed in
Section~\ref{sec:related}.

Most football prediction models describe a team's current strength with one number, such as its current Elo rating, or with a small attack and defence summary. However, an Elo rating is part of a time series. Its recent movement can also be useful. For example, two teams may have the same current rating, but one may be improving while the other is declining. A direct way to use this information is to include several lagged Elo differences in the model. However, these lagged differences are usually highly correlated. Because of this, adding them directly can make the model unstable and may not improve prediction. This creates a dimensionality problem, which has received limited attention in football forecasting.

We address this problem using sufficient dimension reduction. For each match, we first create a vector of lagged Elo differences. We then use categorical SDR, SIR and SAVE, to reduce this vector to a small number of informative directions. These directions are used in a Poisson goal model, which gives win, draw, loss, and scoreline probabilities. We compare these models with logistic regression, standard Poisson regression, time-series models, and machine-learning baselines. The comparison is based on the 2018 and 2022 World Cups, using the ranked probability score. Finally, we apply the best model to forecast the 2026 tournament. The main result is that the two-direction SDR model outperforms the other models.

\section{Related Work}
\label{sec:related}
This study connects three areas: statistical models for football scores, rating systems for measuring team strength, and sufficient dimension reduction.

\paragraph{Goal-based statistical models.}
A common approach in football prediction is to model the number of goals scored by each team. \citep{maher1982} modelled home and away goals as two independent Poisson variables, using attack and defence parameters for each team. \citep{dixon1997} improved this model by adding a correction for low scores and giving more weight to recent matches. Later, \citep{karlis2003} studied the bivariate Poisson model and showed how goal probabilities can be converted into win, draw, and loss probabilities. \citep{rue2000} also used a dynamic version of the Poisson model, where team strengths can change over time.
In this paper, we also use a Poisson goal model. However, instead of estimating many team-specific attack and defence parameters, we use a smaller strength summary based on Elo. This is more suitable for international football, because many national teams have only a limited number of matches against each other.

\paragraph{Rating systems and Elo.}
Another common approach is to describe each team's strength with one rating that changes over time. The Elo system \citep{elo1978}, first developed for chess, updates a team's rating after each match based on the difference between the actual result and the expected result. In football, \citep{hvattum2010} showed that the Elo difference between two teams is one of the most useful variables for predicting match outcomes. \citep{lasek2013} also found that Elo-type ratings perform well compared with other simple rating systems.
For this reason, we use the Elo difference as the main measure of relative team strength. However, we do not only use its current value. We also use its recent history, because the movement of a team's rating may contain useful information for prediction.

\paragraph{Machine learning and tournament forecasting.}
Machine-learning methods have also been used for football prediction. \citep{hubacek2019} showed that gradient-boosted trees can perform well in football betting markets, and \citep{bunker2019} reviewed machine-learning methods in sports prediction and found gradient boosting to be one of the stronger classifiers. For World Cup prediction, \citep{groll2019} used boosted trees together with national-team covariates.
Another useful idea is to combine several forecasts. \citep{bates1969} showed that averaging forecasts can reduce prediction error, and \citep{timmermann2006} reviewed forecast-combination methods in detail. \citep{leitner2010} applied this idea to international football tournaments.
Based on this literature, we include logistic regression, Poisson regression, gradient boosting, and an ensemble model as baselines.

\paragraph{Time-series forecasting of ratings.}
Since Elo ratings change over time, another question is whether forecasting future Elo values can improve match prediction. Some earlier studies have used dynamic models for football outcomes and team strengths, such as \citep{goddard2005} and \citep{crowder2002}.
In this paper, we test this idea using two time-series models. The first is ARIMA, which is a standard linear forecasting model \citep{box2015,hyndman2018}. The second is neural-network autoregression, which can capture nonlinear patterns in the rating series \citep{hyndman2018,zhang1998}. For each team, these models forecast the Elo rating one month ahead. The forecasted Elo difference is then used in the Poisson goal model. Related neural-network approaches for sports prediction have also been used by \citep{loeffelholz2009} and \citep{tax2015}.

\paragraph{Sufficient dimension reduction.}
Sufficient dimension reduction (SDR) seeks a low-dimensional projection of the predictors ($\mathbf{B}^\top$X) such that the conditional distribution of the response given the full predictor vector is preserved, that is, 
\begin{align*}
    Y \mathrel{\perp\!\!\!\perp} X \mid \mathbf{B}^\top X .
\end{align*}
For categorical responses, this framework is closely related to discriminant analysis. \citep{cook2001dimension} show that classical discriminant methods can be interpreted from an SDR perspective. In particular, the population subspace recovered by sliced inverse regression (SIR) is equivalent to the LDA subspace for constructing low-dimensional summary plots. Thus, LDA can be viewed as a mean-based SDR method for classification problems. SAVE provides a complementary SDR approach because it uses class-conditional covariance information and can therefore detect structure not captured by class means alone  \citep{cook1991,cook2000save}. In football analytics, discriminant analysis has been used to separate winning, drawing, and losing teams based on match statistics \citep{lagopenas2010,castellano2012,liu2015}, but it has mostly been used descriptively.
In this paper, we use SIR and SAVE only as supervised dimension-reduction preprocessors for the lagged Elo-difference vector. Because SIR and LDA give the same directions in this whitened categorical setting, we report SIR and do not list LDA as a separate model \citep{cook2001dimension}; the final probabilistic forecasting model is the Poisson goal model fitted to the resulting reduced scores. \citep{zhang2019efficient} note that reduce-then-classify procedures do not necessarily improve prediction. Therefore, we evaluate the reduced-score models empirically using out-of-sample ranked probability scores.


\paragraph{Our contribution.}
This paper makes three main contributions. First, we represent team strength using a short history of lagged Elo differences, instead of using only the current Elo difference. We then reduce this history with categorical SDR and use the reduced scores in a Poisson goal model.
Second, we show that reducing the Elo history to one or two directions improves on the baselines, with the first direction carrying almost all of the gain and the second adding a small refinement to draw probabilities.
Third, we compare the proposed SDR models with logistic regression, Poisson regression, time-series models, machine-learning models, and an ensemble under the same ranked-probability-score evaluation protocol. The comparison is based on the 2018 and 2022 World Cups, and the two primary SDR models, SIR and SAVE with two directions, are then used to forecast the 2026 tournament.
\section{Data}
\label{sec:data}
\subsection{International match results}
The data source is \citep{jurisoo2023}, available at 
\url{https://github.com/martj42/international_results}. The dataset
contains 49{,}257 matches from November 30, 1872, to June 1, 2026.
Variables include date, home and away teams, scores, tournament name,
city, country, and a neutral-venue indicator. Matches with missing
scores (72 records) are excluded. 
Cura\c{c}ao has insufficient match history for reliable estimation of the SDR-based Elo-history projection. Therefore, Cura\c{c}ao is excluded from the estimation of the SDR directions, but it is retained in the 2026 tournament simulation through an Elo-only Poisson fallback model described in Section  \ref{sec:pred2026}. 


\subsection{Elo ratings}
We measure the strength of each team using the Elo rating system
\citep{elo1978,hvattum2010}. Elo gives every team a single rating that
is updated after each match. The size of the update depends on the
match result and on the result that was expected from the two
pre-match ratings. After a match between a home team $h$ and an away
team $a$, the home rating is updated as
\begin{equation}
  R'_{h} = R_{h} + \kappa\,\Gamma\,(S_{h} - E_{h}),
  \qquad
  E_{h} = \frac{1}{1+10^{-(R_{h}+\zeta-R_{a})/400}},
\label{eq:elo}
\end{equation}
where $R_{h}$ and $R_{a}$ are the pre-match ratings of the home and
away teams, $R'_{h}$ is the updated home rating, and $E_{h}$ is the
expected score implied by the rating difference. The away rating is
updated in the same way, so the two changes are equal and opposite.
The actual score is $S_{h}\in\{0,0.5,1\}$, with $1$ for a home win,
$0.5$ for a draw, and $0$ for a home loss. The constant $\zeta$ is the
home advantage: $\zeta=100$ on non-neutral venues and $\zeta=0$ for
matches played at a neutral venue, including all World Cup matches.
All teams start at $R_0=1{,}500$.

The term $\Gamma$ is a goal-difference multiplier. It increases the
rating change when a team wins by a larger margin. Let
$\Delta g_m=|G^{H}_m-G^{A}_m|$ be the absolute goal difference (the
winning margin) of match $m$. Then
\begin{equation}
  \Gamma =
  \begin{cases}
    1, & \Delta g_m \le 1,\\[3pt]
    \dfrac{3}{2}, & \Delta g_m = 2,\\[3pt]
    \dfrac{11+\Delta g_m}{8}, & \Delta g_m \ge 3.
  \end{cases}
\label{eq:gamma}
\end{equation}
A draw or a one-goal win keeps the update unscaled ($\Gamma=1$), a
two-goal win scales it by $3/2$, and every further goal adds $1/8$ to
the multiplier (for example, $\Gamma=7/4$ for a three-goal win and
$\Gamma=2$ for a five-goal win). This is the standard goal-difference
index used in the World Football Elo ratings.

The term $\kappa$ is a match-importance weight. It is larger for more
important matches, so that competitive games change the ratings more
than friendlies. We assign
\begin{equation}
  \kappa =
  \begin{cases}
    60, & \text{FIFA World Cup (final tournament)},\\
    35, & \text{continental championships (e.g.\ UEFA Euro, Copa Am\'erica, AFCON)},\\
    25, & \text{World Cup and continental qualifiers},\\
    20, & \text{friendlies and all other matches}.
  \end{cases}
\label{eq:kappa}
\end{equation}
With this scheme, World Cup results carry the most weight and
friendlies the least.

We use Elo as our main measure of relative team strength for three
reasons: it is simple to compute and fully reproducible, it updates
automatically after every match, and it is among the most accurate of
the simple rating systems for international football
\citep{hvattum2010,lasek2013}.

\subsection{2026 World Cup groups}
The official draw was held on December 5, 2025, in Washington, D.C.
Table~\ref{tab:groups} lists the 12 groups. 

\begin{table}[H]
\centering\footnotesize
\caption{Official 2026 FIFA World Cup group assignments. Source: FIFA draw, December 2025.}
\label{tab:groups}
\renewcommand{\arraystretch}{1.2}
\begin{tabular}{clclcl}
\toprule
\textbf{Grp}&\textbf{Teams}&\textbf{Grp}&\textbf{Teams}&\textbf{Grp}&\textbf{Teams}\\
\midrule
A & Mexico, South Korea,    & B & Canada, Bosnia,       & C & Brazil, Morocco,\\
  & South Africa, Czechia   &   & Qatar, Switzerland    &   & Haiti, Scotland \\
D & USA, Paraguay,          & E & Germany, Ivory Coast, & F & Netherlands, Japan,\\
  & Australia, Turkey       &   & Cura\c{c}ao, Ecuador  &   & Sweden, Tunisia \\
G & Belgium, Egypt,         & H & Spain, Cape Verde,    & I & France, Senegal,\\
  & Iran, New Zealand       &   & Saudi Arabia, Uruguay &   & Iraq, Norway   \\
J & Argentina, Algeria,     & K & Portugal, DR Congo,   & L & England, Croatia,\\
  & Austria, Jordan         &   & Uzbekistan, Colombia  &   & Ghana, Panama  \\
\bottomrule
\end{tabular}
\end{table}

\section{Methodology}
\label{sec:method}

\subsection{Notation}
Table~\ref{tab:notation} collects the notation used throughout the
paper. For each match $m$ between home team $h$ and away team $a$, all
features are computed from data observed strictly before the match
date.

\begin{footnotesize}
\renewcommand{\arraystretch}{1.25}
\begin{longtable}{p{0.20\textwidth} p{0.72\textwidth}}
\caption{Notation used throughout the paper.}
\label{tab:notation}\\
\toprule
\textbf{Symbol} & \textbf{Definition} \\
\midrule
\endfirsthead
\multicolumn{2}{l}{\small\textit{Table~\ref{tab:notation} continued.}}\\[2pt]
\toprule
\textbf{Symbol} & \textbf{Definition} \\
\midrule
\endhead
\midrule
\multicolumn{2}{r}{\small\textit{Continued on next page.}}\\
\endfoot
\bottomrule
\endlastfoot
\multicolumn{2}{l}{\textit{Indices, sets, and dimensions}}\\
$m$ & Match index \\
$h,\;a$ & Home-team and away-team subscripts for match $m$ \\
$n$ & Team index in $R_{n,t}$; also the training sample size when unsubscripted \\
$t$ & Calendar month index; $t_m$ is the month of match $m$;\\
$T$ & The most recent month \\
$c$ & Outcome class treated as a slice in SDR, $c\in\{H,D,A\}$ \\
$r$ & Boosting round, $r=1,\ldots,400$ \\
$C$ & Number of outcome classes, $C=3$ \\
$K$ & Number of lagged monthly Elo differences, $K=6$ \\
$d$ & SDR structural dimension \\
$N$ & Number of Monte Carlo tournament replications, $N=5{,}000$ \\
\midrule
\multicolumn{2}{l}{\textit{Elo ratings}}\\
$R_{n,t}$ & Elo rating of team $n$ in month $t$ \\
$R_{h,m},\;R_{a,m}$ & Pre-match Elo of the home and away team \\
$R'_{h}$ & Post-match (updated) Elo of the home team \\
$R_0$ & Initial Elo rating, $R_0=1{,}500$ \\
$\hat{R}_{n,T+1}$ & One-step-ahead Elo forecast (ARIMA or NNAR) \\
$\kappa$ & Match-importance weight, $\kappa\in\{20,25,35,60\}$ \\
$\Gamma$ & Goal-difference multiplier in the Elo update \\
$\zeta$ & Home-advantage constant ($\zeta=0$ for World Cup matches) \\
$S_{h}$ & Actual match-result score, $S_{h}\in\{0,0.5,1\}$ \\
$E_{h}$ & Elo-expected score of the home team \\
$\Delta_m=R_{h,m}-R_{a,m}$ & Current (lag-0) Elo difference \\
$\hat{\Delta}^{\rm AR}_m,\;\hat{\Delta}^{\rm NNAR}_m$ & Forecasted Elo difference from ARIMA / NNAR \\
$\hat{p}^H_m$ & Elo-implied home win probability, $[1+10^{-(\Delta_m+\zeta)/400}]^{-1}$ \\
\midrule
\multicolumn{2}{l}{\textit{Match-level features}}\\
$N_m\in\{0,1\}$ & Neutral-venue indicator ($N_m=1$ for all WC matches) \\
$\bar{G}^+_h,\;\bar{G}^-_h$ & Home rolling goals scored/conceded (mean of last 6 matches) \\
$\bar{G}^+_a,\;\bar{G}^-_a$ & Away rolling goals scored / conceded \\
$g^+_{n,m-i},\;g^-_{n,m-i}$ & Goals scored/conceded by team $n$ in its $i$-th most recent match before $m$ \\
$\mathrm{pts}_{n,j}\in\{0,1,3\}$ & Points won by team $n$ in its $j$-th most recent match \\
$\bar{F}_{n,m}$ & Rolling form (mean points, last 6) for team $n$ \\
$\bar{F}_m=\bar{F}_{h,m}-\bar{F}_{a,m}$ & Form-points differential \\
$\mathbf{x}_m$ & Lagged Elo-difference feature vector, $\mathbf{x}_m\in\mathbb{R}^K$ \\
$\tilde{\mathbf{x}}_m$ & Whitened lagged Elo-difference feature vector, $\tilde{\mathbf{x}}_m=\widehat{\Sigma}_X^{-1/2}(\mathbf{x}_m-\bar{\mathbf{x}})$\\
$\bar{\mathbf{x}},\widehat{\Sigma}_X$ &  Training mean and covariance matrix of $\mathbf{x}_m$ \\
\midrule
\multicolumn{2}{l}{\textit{Outcome and goal model}}\\
$Y_m\in\{H,D,A\}$ & Observed match outcome \\
$G^H_m,\;G^A_m$ & Actual goals scored by home / away team \\
$g_h,\;g_a$ & Goal-count arguments in the scoreline summation \\
$g_{\max}$ & Maximum goals per team in the summation, $g_{\max}=8$ \\
$\lambda^H_m,\;\lambda^A_m$ & Expected (Poisson-mean) goals, home / away \\
$\hat{\mathbf{p}}_m=(\hat{p}^H_m,\hat{p}^D_m,\hat{p}^A_m)$ & Predicted outcome-probability vector, summing to one \\
$\mathbf{1}[\cdot]$ & Indicator function \\
\midrule
\multicolumn{2}{l}{\textit{Logistic regression (M1, M2)}}\\
$\alpha_k$ & Class-$k$ intercept \\
$\beta_k$ & Class-$k$ Elo-difference slope (M1) \\
$\gamma_k$ & Class-$k$ neutral-venue slope (M1) \\
$\boldsymbol{\beta}_k$ & Class-$k$ slope vector (M2; 8 components) \\
\midrule
\multicolumn{2}{l}{\textit{Poisson double regression (M3)}}\\
$\mu^H,\;\mu^A$ & Baseline log-goal intercepts, home / away \\
$\xi$ & Elo-difference coefficient (single-direction models) \\
$\xi_j$ & Coefficient on the $j$-th SDR score $z_{m,j}$ (M8--M11) \\
$\delta^H,\;\delta^A$ & Neutral-venue coefficients, home / away \\
$\eta_1,\ldots,\eta_4$ & Rolling-goal (form) coefficients \\
\midrule
\multicolumn{2}{l}{\textit{Time-series strength forecasting (M4 ARIMA, M5 NNAR)}}\\
$B$ & Backshift operator, $B\,R_{n,t}=R_{n,t-1}$ \\
$\phi(B);\ \phi_1,\ldots,\phi_p$ & AR polynomial and coefficients \\
$\theta(B);\ \theta_1,\ldots,\theta_q$ & MA polynomial and coefficients \\
$d_{\rm AR}$ & ARIMA degree of differencing \\
$\varepsilon_{n,t}$ & White-noise innovation, $\mathrm{WN}(0,\sigma^2_n)$ \\
$\hat{\ell}$ & Maximised log-likelihood (used in AIC) \\
$P$ & Number of seasonal lags in NNAR (period 12; $P=1$) \\
$k_{\rm h}$ & Number of hidden nodes in NNAR \\
$\sigma(\cdot)$ & Logistic sigmoid activation \\
$w_0,\ldots,w_{k_{\rm h}}$ & NNAR output-layer weights \\
$v_{j0},\,v_{ji}$ & NNAR input-to-hidden weights for node $j$ (bias $v_{j0}$) \\
\midrule
\multicolumn{2}{l}{\textit{Gradient boosting (M6)}}\\
$F^{(r)}_k(\mathbf{x}_m)$ & Additive raw score for class $k$ after round $r$ \\
$h^{(r)}_k$ & $r$-th regression tree for class $k$ \\
$\nu$ & Learning rate (shrinkage), $\nu=0.05$ \\
$\mathcal{L}$ & Multinomial log-loss objective \\
$y_m$ & Observed outcome label of match $m$ \\
\midrule
\multicolumn{2}{l}{\textit{Categorical sufficient dimension reduction (M8--M11)}}\\
$\mathbf{M}$ & Generic SDR kernel matrix, $\mathbf{M}\in\mathbb{R}^{K\times K}$ \\
$\bar{\mathbf{x}}_c$ & Class-conditional mean of $\tilde{\mathbf{x}}_m$ \\
$\widehat{\boldsymbol{\Sigma}}_c$ & Within-class covariance matrix \\
$n_c$ & Number of matches in class $c$ ($n_H,n_D,n_A$) \\
$\hat{\pi}_c=n_c/n$ & Class proportion \\
$\mathbf{I}_K$ & $K\times K$ identity matrix \\
$\hat{\lambda}_j$ & $j$-th eigenvalue of the SDR kernel \\
$\mathbf{B}=[\hat{\boldsymbol{\beta}}_1,\ldots,\hat{\boldsymbol{\beta}}_d]$ & $K\times d$ projection matrix \\
$\mathbf{z}_m=\mathbf{B}^\top\tilde{\mathbf{x}}_m$ & SDR score vector; $z_{m,j}$ is the $j$-th score \\
\midrule
\multicolumn{2}{l}{\textit{Evaluation and tournament simulation}}\\
$\mathrm{RPS}_m$ & Per-match ranked probability score \\
$\overline{\mathrm{RPS}}$ & Mean RPS over the test set \\
$\widehat{\mathrm{xPts}}_t$ & Projected group points for team $t$ \\
\end{longtable}
\end{footnotesize}

\subsection{M1: Elo-Logistic Regression}
\label{sec:m1}
Our baseline is M1, a multinomial logistic regression that
predicts the match outcome directly from a single strength summary,
the Elo difference. \citep{hvattum2010} compared many candidate
predictors for international football and found that the Elo rating
difference was the most informative single variable. M1 therefore
keeps only this predictor, together with a neutral-venue indicator,
and serves as the simplest model against which every later model is
measured. For a match $m$ with Elo difference
$\Delta_m=R_{h,m}-R_{a,m}$ and neutral-venue indicator $N_m$, the
probability of each outcome $k\in\{H,D,A\}$ is given by the
multinomial logit \citep{agresti2002}:
\begin{equation*}
  P(Y_m=k\mid\Delta_m,N_m) =
  \frac{\exp(\alpha_k+\beta_k\Delta_m+\gamma_k N_m)}
       {\sum_{j\in\{H,D,A\}}\exp(\alpha_j+\beta_j\Delta_m+\gamma_j N_m)},
  \qquad k\in\{H,D,A\}.
\label{eq:m1}
\end{equation*}
Here $\alpha_k$ is the class-$k$ intercept, $\beta_k$ measures how the
Elo difference shifts the log-odds of outcome $k$, and $\gamma_k$
measures the effect of a neutral venue. There are nine coefficients in
all, three per class. The model assumes that the log-odds of each
outcome are linear in the Elo difference and additive in the
neutral-venue indicator, and that matches are conditionally
independent given $(\Delta_m,N_m)$. The
coefficients are estimated by maximum likelihood over the
$n=20{,}775$ training matches:
\begin{equation*}
  \hat{\boldsymbol{\theta}} = \argmax_{\boldsymbol{\theta}}
  \sum_{m=1}^{n}\log P(Y_m\mid\Delta_m,N_m;\,\boldsymbol{\theta}).
  \label{eq:m1_mle}
\end{equation*}
Where $\boldsymbol{\theta}=(\alpha_{D},\alpha_{A},\beta_{D},\beta_{A},\gamma_{D},\gamma_{A})$. Because the three probabilities must sum to one, one class is fixed
for identification. We set $\alpha_H=\beta_H=\gamma_H=0$, which leaves
six free parameters, with estimates
\begin{equation*}
  \begin{pmatrix}\hat{\alpha}_H\\\hat{\alpha}_D\\\hat{\alpha}_A\end{pmatrix}
  =\begin{pmatrix}0\\-0.5932\\-0.7781\end{pmatrix},\quad
  \begin{pmatrix}\hat{\beta}_H\\\hat{\beta}_D\\\hat{\beta}_A\end{pmatrix}
  =\begin{pmatrix}0\\-0.00328\\-0.00629\end{pmatrix},\quad
  \begin{pmatrix}\hat{\gamma}_H\\\hat{\gamma}_D\\\hat{\gamma}_A\end{pmatrix}
  =\begin{pmatrix}0\\+0.262\\+0.670\end{pmatrix}.
  \label{eq:m1_estimates}
\end{equation*}
To show how a prediction is generated, consider an illustrative
neutral-venue match ($N_m=1$) between Spain and Morocco with Elo
ratings $R_{\text{Spain}}=2{,}123$ and $R_{\text{Morocco}}=1{,}923$,
so $\Delta_m = 200$. The linear predictors are
$\hat{u}_D = -0.5932 + (-0.00328)(200) + (0.262)(1) = -0.985$ and
$\hat{u}_A = -0.7781 + (-0.00629)(200) + (0.670)(1) = -1.364$ (with
$\hat u_H=0$). Exponentiating and normalising gives
$P(\text{Spain})=61.4\%$, $P(\text{Draw})=22.9\%$, and
$P(\text{Morocco})=15.7\%$. The same coefficients apply to every
match; only $\Delta_m$ and $N_m$ change between fixtures. Like all
later models, M1 is assessed out of sample on the 2018 and 2022 World
Cups using the ranked probability score (Section~\ref{sec:RSL}).

\subsection{M2: Full Logistic Regression}
\label{sec:m2}
M2 keeps the multinomial framework of M1, but lets us test whether recent team form adds information beyond the Elo difference
\citep{groll2019}. It uses the full feature vector
$\mathbf{x}_m=(\Delta_m,\hat{p}^H_m,N_m,\bar{F}_m,\bar{G}^+_h,
\bar{G}^-_h,\bar{G}^+_a,\bar{G}^-_a)^\top$:
\begin{equation*}
  P(Y_m=k) =
  \frac{\exp(\alpha_k+\boldsymbol{\beta}_k^\top\mathbf{x}_m)}
  {\sum_{j}\exp(\alpha_j+\boldsymbol{\beta}_j^\top\mathbf{x}_m)},
\label{eq:m2}
\end{equation*}
where $\boldsymbol{\beta}_k\in\mathbb{R}^8$ is a vector of eight slope
coefficients. Fixing $\alpha_H=0$ and $\boldsymbol{\beta}_H=\mathbf{0}$
leaves 18 free parameters, estimated by maximum likelihood over
$n=20{,}775$ matches. The rolling goals scored and conceded for the home team are
\begin{equation*}
  \bar{G}^+_h = \frac{1}{6}\sum_{i=1}^{6} g^+_{h,m-i},
  \qquad
  \bar{G}^-_h = \frac{1}{6}\sum_{i=1}^{6} g^-_{h,m-i},
\end{equation*}
and symmetrically for the away team. The rolling form for team $n$ is
$\bar{F}_{n,m} = \tfrac{1}{6}\sum_{j=1}^{6}\mathrm{pts}_{n,m-j}$, and
the form differential entering the model is
$\bar{F}_m = \bar{F}_{h,m} - \bar{F}_{a,m}$. These features capture
attacking and defensive form, and winning momentum, independently of
the slowly updating Elo rating \citep{dixon1997}.

\subsection{M3: Poisson Double Regression}
\label{sec:m3}
Instead of modelling the win, draw, and loss outcome directly, M3 models the number of goals scored by each team. This gives richer output, because the model can produce both scoreline probabilities and match-outcome probabilities.
This approach follows the Poisson football-modeling literature. \citep{maher1982} modelled home and away goals as independent Poisson variables with team-level attack and defence parameters. \citep{dixon1997} later added time-decay weights and a correction for low scores. \citep{karlis2003} discussed the bivariate Poisson framework and the conversion from scoreline probabilities to win, draw, and loss probabilities. \citep{rue2000} used a related likelihood in a Bayesian dynamic model. Our model follows this general idea, but we do not estimate separate attack and defence parameters for every national team. Instead, we use the Elo difference $\Delta_m$ as the main strength summary \citep{hvattum2010}. This is more stable for international football, because many national-team pairs have only a small number of head-to-head matches. Home and away goals are modelled as conditionally independent Poisson variables:
\begin{align}
  G^H_m &\sim \mathrm{Poisson}(\lambda^H_m),\quad
  \log\lambda^H_m = \mu^H + \xi\Delta_m + \delta^H N_m
    + \eta_1\bar{G}^+_h + \eta_2\bar{G}^-_a,
\label{eq:m3h}\\
  G^A_m &\sim \mathrm{Poisson}(\lambda^A_m),\quad
  \log\lambda^A_m = \mu^A - \xi\Delta_m + \delta^A N_m
    + \eta_3\bar{G}^+_a + \eta_4\bar{G}^-_h.
\label{eq:m3a}
\end{align}
Fitting M3 on the full pre-2026 training dataset gives, $\hat{\mu}^H=0.035$, $\hat{\xi}=0.00146$,
$\hat{\delta}^H=-0.085$, $\hat{\delta}^A=+0.263$, $\hat{\mu}^A=-0.442$,
$\hat{\eta}_1=0.090$, $\hat{\eta}_2=0.157$, $\hat{\eta}_3=0.108$, and
$\hat{\eta}_4=0.167$. Since $G^H_m$ and $G^A_m$ are conditionally
independent, their joint mass factorises.   Match-outcome probabilities are obtained by summing over all scorelines consistent
with each outcome.  Because the scoreline summation is truncated at $g_{\max}=8$,  we first compute the three unnormalized probabilities on the same truncated grid as  
\begin{equation}
\begin{aligned}
  P(Y_m = H) &= \sum_{g_h=0}^{g_{\max}}\sum_{g_a=0}^{g_h-1}
    P(G^H_m=g_h)\,P(G^A_m=g_a), \\[4pt]
  P(Y_m = D) &= \sum_{g=0}^{g_{\max}}
    P(G^H_m=g)\,P(G^A_m=g), \\[4pt]
  P(Y_m = A) &= \sum_{g_a=0}^{g_{\max}}\sum_{g_h=0}^{g_a-1}
    P(G^H_m=g_h)\,P(G^A_m=g_a). 
\end{aligned}
\label{eq:pois_probs}
\end{equation}
The final win, draw, and loss probabilities are then obtained by normalizing:
\begin{align*}
    \widehat{P}(Y_m=k)=\frac{P(Y_m = k)}{P(Y_m = H)+ P(Y_m = D)+ P(Y_m = A)}, ~~~~~~ k\in\{ H, D, A\}.
\end{align*}
We use $g_{\max}=8$,  the maximum number of goals per team considered in the summation. The omitted probability mass outside this grid is negligible for the fitted goal rates. 
For Spain vs.\ Morocco with $\hat{\lambda}^H=1.68$ and
$\hat{\lambda}^A=0.89$, summing over all scorelines grid and normalizing gives 
$P(Y_m=H)=63.1\%$, $P(Y_m=D)=22.6\%$, $P(Y_m=A)=14.3\%$.
In the 2018 and 2022 backtests, all coefficients are re-estimated using only matches before the corresponding information barrier.
\subsection{M4: ARIMA Poisson}
\label{sec:arima}
M4 is the same Poisson goal model as M3, but with one change. Instead of using each team's current Elo rating, it uses a one-month-ahead forecast of that team's Elo rating. The reason is that Elo ratings change over time. A team whose rating is increasing may be stronger than its current rating suggests, while a team whose rating is decreasing may be weaker. To test this idea, we model each team's monthly Elo series $\{R_{n,t}\}_{t=1}^T$ as a univariate time series and fit an ARIMA$(p,d_{\rm AR},q)$ model \citep{box2015,hyndman2018}. ARIMA is a standard method for one-step-ahead forecasting and has also been used in football-related forecasting studies \citep{goddard2005,crowder2002}:

\begin{equation*}
  \phi(B)(1-B)^{d_{\rm AR}}\,R_{n,t} =
  c + \theta(B)\,\varepsilon_{n,t},
  \qquad \varepsilon_{n,t} \sim \mathrm{WN}(0,\,\sigma^2_n).
\label{eq:arima}
\end{equation*}
Here $B$ is the backshift operator. The parameters $p$, $d_{\rm AR}$, and $q$ control the ARIMA structure: $p$ is the number of autoregressive lags, $d_{\rm AR}$ is the differencing order, and $q$ is the number of moving-average terms. The term $c$ represents the drift constant; the polynomials are
$\phi(B)=1-\phi_1 B-\cdots-\phi_p B^p$ and
$\theta(B)=1+\theta_1 B+\cdots+\theta_q B^q$. Rather than impose one
order on every team, we select $(p,d_{\rm AR},q)$ per team by
minimising the Akaike Information Criterion \citep{akaike1974},
$\mathrm{AIC} = -2\hat{\ell} + 2(p+q+1)$, using \texttt{auto.arima()}
in R \citep{hyndman2018}, which searches $p,q \leq 5$ and
$d_{\rm AR}\leq 2$. Table~\ref{tab:arima_orders} reports the orders
chosen.
 
\begin{table}[H]
\centering\small
\caption{Distribution of ARIMA$(p,d_{\rm AR},q)$ orders selected by
  AIC across the 47 WC-qualified teams (monthly Elo series, January
  2000 -- March 2026).}
\label{tab:arima_orders}
\renewcommand{\arraystretch}{1.3}
\begin{tabular}{cccr}
\toprule
$p$ & $d_{\rm AR}$ & $q$ & \textbf{Number of teams} \\
\midrule
0 & 1 & 0 & 30 \\
1 & 0 & 0 &  3 \\
1 & 1 & 1 &  3 \\
0 & 1 & 1 &  2 \\
0 & 1 & 3 &  2 \\
1 & 1 & 0 &  2 \\
Other &  &   &  5 \\
\bottomrule
\end{tabular}
\end{table}

The forecasted Elo difference is defined as \[ \hat{\Delta}^{\rm AR}_m = \hat{R}_{h,T+1}^{\rm AR} - \hat{R}_{a,T+1}^{\rm AR}. \] This value replaces the current Elo difference $\Delta_m$ in equations~\eqref{eq:m3h}--\eqref{eq:m3a}. Within each backtest, the same Poisson specification as M3 is fitted on the training data, but the current Elo difference is replaced by the ARIMA-forecasted Elo difference, and the final win, draw, and loss probabilities are obtained as in equation~\eqref{eq:pois_probs}. In practice, the ARIMA forecast is very close to the current Elo rating for most teams. Therefore, M4 gives almost the same results as M3 (combined RPS $0.213$ vs.\ $0.212$; Table~\ref{tab:backtest_all}). This suggests that forecasting the Elo level one step ahead does not add useful predictive information in this setting. The useful trajectory information enters the model in a different way: instead of forecasting the next Elo level, we use the vector of lagged Elo differences and reduce it with the SDR methods described in Section~\ref{sec:sdr_preamble}.

\subsection{M5: Neural Network Autoregression (NNAR) Poisson}
\label{sec:nnar}
NNAR was used by \citep{hyndman2018} through the
\texttt{nnetar()} function in R. It has also been used in sports
prediction. For example, \citep{loeffelholz2009} applied neural
networks to NBA performance series, and \citep{tax2015} used neural
networks for Dutch club football prediction.
We use NNAR to test whether nonlinear changes in Elo ratings contain
useful information. For example, a team may improve quickly after a new
coach, or decline after a poor qualifying campaign. These patterns may
not be fully captured by a linear ARIMA model \citep{zhang1998}.
NNAR$(p,P,k_{\rm h})$ is a feedforward neural network with one hidden
layer. It uses $p$ non-seasonal lagged values and $P$ seasonal lagged
values as inputs, with $k_{\rm h}$ hidden nodes:

\begin{equation*}
  \hat{R}_{n,T+1} = w_0 +
  \sum_{j=1}^{k_{\rm h}} w_j \,
  \sigma\!\left(
    v_{j0} +
    \sum_{i=1}^{p} v_{ji}\,R_{n,T+1-i} +
    \sum_{s=1}^{P} v_{j,p+s}\,R_{n,T+1-12s}
  \right),
  \label{eq:nnar}
\end{equation*}
where $\sigma(z)=(1+e^{-z})^{-1}$ is the logistic sigmoid,
$w_0,\ldots,w_{k_{\rm h}}$ are the output-layer weights, and
$v_{j0},v_{j1},\ldots$ are the input-to-hidden weights for node $j$.
All weights are estimated by minimising the one-step-ahead mean
squared error \citep{hyndman2018}. The lag order $p$ is taken from
\texttt{auto.arima()} and the number of hidden nodes is
$k_{\rm h} = \lfloor (p + P + 1)/2 \rfloor$, with $P=1$ seasonal lag
(lag 12) for all teams. The dominant architecture is NNAR$(1,1,2)$,
selected for $81\%$ of teams. Within each backtest, the NNAR-forecasted Elo difference, $\hat{\Delta}^{\rm NNAR}_m$, replaces the current Elo difference in the same Poisson double-regression framework.

\subsection{M6: XGBoost}
\label{sec:xgb}
\citep{chen2016} introduced XGBoost as a scalable and regularised
version of gradient-boosted trees. In football prediction,
\citep{hubacek2019} showed that XGBoost can perform well in betting
markets, while \citep{groll2019} used boosted trees with team-level
covariates for World Cup forecasting. More generally,
\citep{bunker2019} found gradient boosting to be one of the stronger
classifiers in sports prediction.
We include XGBoost as a machine-learning baseline. The model starts
with a zero raw score for each match and each outcome,
\[
F^{(0)}_k(\mathbf{x}_m)=0,
\qquad k\in\{H,D,A\}.
\]
It then adds one regression tree at each boosting round. In our implementation, we use 400 boosting rounds:
\begin{equation*}
  F^{(r)}_k(\mathbf{x}_m) =
  F^{(r-1)}_k(\mathbf{x}_m) + \nu\, h^{(r)}_k(\mathbf{x}_m),
  \label{eq:xgb_update}
\end{equation*}
where each tree $h^{(r)}_k$ has maximum depth 4 (at most
$2^4=16$ leaves) and the learning rate $\nu=0.05$ scales each tree's
contribution so the model learns slowly across rounds rather than
overfitting in a few large steps \citep{friedman2001}. After 400
rounds the three raw scores are mapped to probabilities by the
softmax,
\begin{equation*}
  P(Y_m = k \mid \mathbf{x}_m) =
  \frac{e^{F^{(400)}_k(\mathbf{x}_m)}}
       {\sum_{j\in\{H,D,A\}} e^{F^{(400)}_j(\mathbf{x}_m)}},
  \label{eq:xgb_softmax}
\end{equation*}
and each tree is chosen to minimise the multinomial log-loss
$\mathcal{L} = -\tfrac{1}{n}\sum_{m=1}^{n}
\log P(Y_m = y_m \mid \mathbf{x}_m)$, optimised by a second-order
Taylor expansion that makes the optimal tree structure tractable
\citep{chen2016}.

\subsection{M7: Ensemble}
\label{sec:ens}
An ensemble combines the predictions of several models, usually by averaging their predicted probabilities. This can improve prediction because different models may make different types of errors. Forecast averaging has a long history in statistics \citep{bates1969,timmermann2006}, and it has also been used in international football tournaments \citep{leitner2010}.
Our ensemble model, M7, is a simple average of the outcome probabilities from M1, M2, M3, and M6.

\subsection{Categorical Sufficient Dimension Reduction}
\label{sec:sdr_preamble}
Models M8--M11 share a common framework: they apply categorical sufficient dimension reduction (SDR) to a vector of lagged Elo differences and feed the resulting low-dimensional projection into the same Poisson double regression used in M3.  The use of SIR and SAVE here is as supervised SDR preprocessing, not as a stand-alone classifier.  For each match $m$ between
home team $h$ and away team $a$, played in calendar month $t_m$,
define the $K$-dimensional lagged Elo-difference vector
\begin{equation*}
  \mathbf{x}_m =
  \bigl(
    R_{h,t_m}-R_{a,t_m},\;
    R_{h,t_m-1}-R_{a,t_m-1},\;
    \ldots,\;
    R_{h,t_m-K+1}-R_{a,t_m-K+1}
  \bigr)^\top \in \mathbb{R}^K,
\label{eq:sdr_feat}
\end{equation*}
where $R_{n,t}$ is team $n$'s Elo rating in month $t$. Models M1--M3
use only the first component ($\Delta_m$); M8--M11 use all $K=6$ components.
For the SDR methods, we whiten the lagged Elo-difference vector using the training mean and covariance matrix $\tilde{\mathbf{x}}_m=\widehat{\Sigma}_X^{-1/2}(\mathbf{x}_m-\bar{\mathbf{x}})$, so that the transformed features have training mean zero and training covariance approximately equal to $\mathbf{I}_K$. All SIR and SAVE kernels are computed on $\tilde{\mathbf{x}}_m$.
Each method produces a kernel matrix
$\mathbf{M}\in\mathbb{R}^{K\times K}$ whose top $d$ eigenvectors
\begin{equation*}
  \mathbf{M}\,\hat{\boldsymbol{\beta}}_j =
  \hat{\lambda}_j\,\hat{\boldsymbol{\beta}}_j,
  \qquad j=1,\ldots,d,
  \qquad \hat{\lambda}_1 \geq \hat{\lambda}_2 \geq \cdots,
\label{eq:sdr_eig}
\end{equation*}
form the projection matrix
$\mathbf{B}=[\hat{\boldsymbol{\beta}}_1,\ldots,
\hat{\boldsymbol{\beta}}_d]\in\mathbb{R}^{K\times d}$, giving the score
vector
\begin{equation*}
  \mathbf{z}_m = \mathbf{B}^\top\tilde{\mathbf{x}}_m \in \mathbb{R}^{d}.
\label{eq:sdr_score}
\end{equation*}
The scores $z_{m,1},\ldots,z_{m,d}$ replace $\Delta_m$ as strength inputs in the Poisson double regression:
\begin{align}
  \log\lambda^H_m &= \mu^H + \sum_{j=1}^{d}\xi_j\,z_{m,j}
    + \delta^H N_m + \eta_1\bar{G}^+_h + \eta_2\bar{G}^-_a,
\label{eq:sdr_pois_h}\\
  \log\lambda^A_m &= \mu^A - \sum_{j=1}^{d}\xi_j\,z_{m,j}
    + \delta^A N_m + \eta_3\bar{G}^+_a + \eta_4\bar{G}^-_h.
\label{eq:sdr_pois_a}
\end{align}
The sign convention keeps the interpretation simple: a higher SDR score
for the home team increases the expected number of home goals and
decreases the expected number of away goals. The final win, draw, and
loss probabilities are then obtained as in equation~\eqref{eq:pois_probs}.
All SDR kernels and Poisson coefficients are estimated using
$n=2{,}756$ training matches involving World-Cup-qualified teams from
January 2010 to the relevant World Cup information barrier. The sample
contains $1{,}301$ home wins ($47.2\%$), $656$ draws ($23.8\%$), and
$799$ away wins ($29.0\%$).
The class means also behave as expected. The mean lag-0 Elo difference
is positive for home wins, close to zero for draws, and negative for
away wins:
\[
\bar{x}_0^H = +160.9, \qquad
\bar{x}_0^D = -9.4, \qquad
\bar{x}_0^A = -184.4.
\]
This shows that the current Elo difference already separates the three
outcome classes in the expected direction.
We use Poisson double regression as the response model for M8--M11 for
three reasons. First, goals are naturally modelled as count data, and
Poisson models are standard in football prediction
\citep{maher1982,dixon1997,karlis2003}. Second, modelling the two goal
rates, $\lambda^H_m$ and $\lambda^A_m$, gives a structured way to
regularise the win, draw, and loss probabilities. Third, the model also
produces scoreline probabilities, which are needed for goal-difference
tiebreaking in the tournament simulation.
\subsection{M8 and M9: SIR Poisson}
\label{sec:m8}
Sliced inverse regression \citep{li1991} reduces a categorical
response by treating its classes as slices \citep{cook2001dimension};
here the slices are the three outcome classes $c\in\{H,D,A\}$, and SIR
searches for directions along which the class means of the (whitened)
feature separate. Let
$\mathbf{z}=\boldsymbol{\Sigma}_X^{-1/2}(\mathbf{x}-\boldsymbol{\mu}_X)$
be the population standardised feature, with $\boldsymbol{\mu}_X$ and
$\boldsymbol{\Sigma}_X$ the marginal mean and covariance of
$\mathbf{x}$, class-conditional mean
$\mathbf{m}_c=\mathrm{E}[\mathbf{z}\mid Y=c]$, and class probability
$\pi_c=\Pr(Y=c)$. The population \emph{candidate matrix} for SIR is the
covariance of the inverse-regression curve,
\begin{equation*}
  \mathbf{M}_{\mathrm{SIR}}
  = \mathrm{Cov}\!\bigl(\mathrm{E}[\mathbf{z}\mid Y]\bigr)
  = \sum_{c\in\{H,D,A\}} \pi_c\,\mathbf{m}_c\,\mathbf{m}_c^{\top},
  \label{eq:sir_pop}
\end{equation*}
whose column space is contained in the central subspace
\citep{li1991}. The sample version replaces each population quantity by
its training estimate---$\hat{\pi}_c=n_c/n$, the whitened feature
$\tilde{\mathbf{x}}_m=\widehat{\boldsymbol{\Sigma}}_X^{-1/2}
(\mathbf{x}_m-\bar{\mathbf{x}})$, and the class mean
$\bar{\mathbf{x}}_c=n_c^{-1}\sum_{m:\,Y_m=c}\tilde{\mathbf{x}}_m$---giving
\begin{equation*}
  \widehat{\mathbf{M}}_{\mathrm{SIR}}
  = \sum_{c\in\{H,D,A\}} \hat{\pi}_c\,
    \bar{\mathbf{x}}_c\,\bar{\mathbf{x}}_c^{\top}.
  \label{eq:sir_kernel}
\end{equation*}
The top $d$ eigenvectors of $\widehat{\mathbf{M}}_{\mathrm{SIR}}$ form the
projection matrix $\mathbf{B}$ in \eqref{eq:sdr_eig}: model~M8 uses
$d=1$ and model~M9 uses $d=2$. Because the feature has been whitened,
 the SIR directions coincide with the Fisher linear-discriminant directions; we
therefore report SIR and do not list LDA as a separate model
\citep{cook2001dimension}.

With only $C=3$ classes, the SIR kernel \eqref{eq:sir_kernel} is the
covariance of the class-conditional means, $\mathrm{var}(E[\mathbf{x}\mid Y])$,
whose rank is at most $C-1=2$ because the $C$ centred class means
satisfy one linear constraint. SIR can therefore
recover at most $C-1$ directions
\citep{li1991,cook2001dimension}, the same bound that restricts
Fisher discriminant analysis to $C-1$ discriminant functions. This is strictly tighter than the general SDR bound
$d\le K=6$.

\subsection{M10 and M11: SAVE Poisson}
\label{sec:m9}
Sliced average variance estimation \citep{cook1991,cook2000save} uses
the class-conditional \emph{covariances} rather than the class means.
It can therefore detect directions along which the outcome classes
differ in spread, even where their means coincide---exactly the kind
of structure that separates draws (a low-variance, evenly matched
regime) from decisive results. With
$\boldsymbol{\Sigma}_c=\mathrm{Cov}(\mathbf{z}\mid Y=c)$ the population
within-class covariance of the standardised feature, the population
candidate matrix for SAVE is
\begin{equation*}
  \mathbf{M}_{\mathrm{SAVE}}
  = \sum_{c\in\{H,D,A\}} \pi_c\,
    \bigl(\mathbf{I}_K-\boldsymbol{\Sigma}_c\bigr)^{2},
  \label{eq:save_pop}
\end{equation*}
which is non-zero along directions where a class covariance departs
from the identity, so its column space captures both mean- and
variance-separation directions of the central subspace
\citep{cook1991,cook2000save}. Replacing $\pi_c$ and
$\boldsymbol{\Sigma}_c$ by their training estimates gives the sample
version
\begin{equation*}
  \widehat{\mathbf{M}}_{\mathrm{SAVE}}
  = \sum_{c\in\{H,D,A\}} \hat{\pi}_c
    \bigl(\mathbf{I}_K-\widehat{\boldsymbol{\Sigma}}_c\bigr)^{2},
  \qquad
  \widehat{\boldsymbol{\Sigma}}_c
  = \frac{1}{n_c-1}\sum_{m:\,Y_m=c}
    (\tilde{\mathbf{x}}_m-\bar{\mathbf{x}}_c)
    (\tilde{\mathbf{x}}_m-\bar{\mathbf{x}}_c)^{\top}.
  \label{eq:save_kernel}
\end{equation*}
A class that is more concentrated than the pooled distribution
($\widehat{\boldsymbol{\Sigma}}_c\neq\mathbf{I}_K$) contributes to
$\widehat{\mathbf{M}}_{\mathrm{SAVE}}$ through both its mean and its covariance,
so SAVE recovers mean-separation \emph{and} variance-separation
directions. The top $d$ eigenvectors of
$\widehat{\mathbf{M}}_{\mathrm{SAVE}}$ again form $\mathbf{B}$: model~M10
uses $d=1$ and model~M11 uses $d=2$.

Models M8--M11 thus span the natural ladder of categorical SDR: a
mean-based reduction (SIR) and a second-moment reduction (SAVE), each at one or two directions. 
We carry \textbf{M9 (SIR, $d=2$)} and \textbf{M11, (SAVE, $d=2)$} forward as the primary forecasting
model; the out-of-sample evaluation in Section~\ref{sec:RSL} confirms
that it attains the lowest-ranked probability score, with M8 and M10 close behind. 
\section{Results}
\label{sec:RSL}
\subsection{Evaluation}
All models are evaluated using two out-of-sample backtests: the 2018
and 2022 FIFA World Cups. For each tournament, we set an information
barrier at the first match of the tournament. The models are fitted
only on matches played before this barrier and are then evaluated on
the World Cup matches after it. No tournament matches are used for
model fitting, feature construction, or hyperparameter selection.
For 2018, the information barrier is June 14, 2018, the opening match
between Russia and Saudi Arabia, giving $n_{2018}=64$ test matches.
For 2022, the barrier is November 20, 2022, the opening match between
Qatar and Ecuador, giving $n_{2022}=64$ test matches. The combined
test set therefore contains $n=128$ matches.
In the World Cup backtests, matches are treated as neutral-site
matches in the model. Therefore, the home and away labels are mainly
used to define the ordering of the two teams, while the neutral-venue
terms $\delta^H$ and $\delta^A$ account for the absence of a standard
home advantage.

For M8--M11, we need the lagged Elo-difference vectors defined in
equation~\eqref{eq:sdr_feat}. Therefore, these models are trained only
on matches involving World-Cup-qualified teams from January 2010 to the
relevant information barrier. This gives $n=2{,}756$ matches for the
2018 barrier and $n=4{,}179$ matches for the 2022 barrier.
For M1--M7, we use the larger match dataset from January 2000 to the
same barrier. All model parameters are re-estimated separately for
each World Cup backtest, so the 2018 and 2022 evaluations are fully
out of sample.

The primary evaluation criterion is the Ranked Probability Score
\citep{epstein1969,gneiting2007}, which for the ordered three-class
problem $H<D<A$ simplifies to
\begin{equation*}
  \mathrm{RPS}_m = \frac{1}{2}\Bigl[
  \bigl(\hat{p}^H_m - \mathbf{1}[Y_m=H]\bigr)^2 +
  \bigl(\hat{p}^H_m+\hat{p}^D_m - \mathbf{1}[Y_m\in\{H,D\}]\bigr)^2
  \Bigr],
\label{eq:rps3}
\end{equation*}
with $\overline{\mathrm{RPS}}=n^{-1}\sum_m \mathrm{RPS}_m$. RPS is a scoring rule that evaluates the full probability distribution, not only the most likely outcome; lower is better.   For a
uniform forecast $(1/3,1/3,1/3)$, the RPS equals $5/18$ for a home or away win and $1/9$ for a draw; if the three outcomes are equally likely, its expected RPS is $2/9$.   
We also report accuracy
(fraction of matches where $\arg\max_k \hat{p}^k_m = Y_m$).
Table~\ref{tab:backtest_all} reports all results.
\begin{table}[H]
\centering\small
\caption{Backtest performance of all models on the 2018 WC ($n=64$),
  2022 WC ($n=64$), and combined ($n=128$) test sets. The categorical
  SDR models (M8--M11) use $K=6$ lagged Elo differences whitened.  M9 (SIR, $d=2$) and  M11 (SAVE, $d=2$) are used to predict 2026 world cup.}
\label{tab:backtest_all}
\renewcommand{\arraystretch}{1.3}
\begin{tabular}{lcccccc}
\toprule
& \multicolumn{2}{c}{\textbf{2018 WC}}
& \multicolumn{2}{c}{\textbf{2022 WC}}
& \multicolumn{2}{c}{\textbf{Combined}}\\
\cmidrule(lr){2-3}\cmidrule(lr){4-5}\cmidrule(lr){6-7}
\textbf{Model}
& RPS$\downarrow$ & Acc$\uparrow$
& RPS$\downarrow$ & Acc$\uparrow$
& RPS$\downarrow$ & Acc$\uparrow$ \\
\midrule
\multicolumn{7}{l}{\textit{Baseline and time-series models}} \\
M1: Elo-Logistic         & 0.218 & 0.516 & 0.220 & 0.500 & 0.219 & 0.508 \\
M2: Full Logistic        & 0.215 & 0.531 & 0.218 & 0.516 & 0.217 & 0.523 \\
M3: Poisson (current Elo)& 0.210 & 0.531 & 0.215 & 0.500 & 0.212 & 0.516 \\
M4: ARIMA--Poisson       & 0.211 & 0.531 & 0.215 & 0.500 & 0.213 & 0.516 \\
M5: NNAR--Poisson        & 0.212 & 0.531 & 0.216 & 0.500 & 0.214 & 0.516 \\
M6: XGBoost              & 0.213 & 0.547 & 0.217 & 0.516 & 0.215 & 0.531 \\
M7: Ensemble (M1--M3, M6)& 0.210 & 0.547 & 0.209 & 0.547 & 0.209 & 0.547 \\
\midrule
\multicolumn{7}{l}{\textit{Categorical SDR Poisson models}} \\
M8:  SIR (LDA) ($d=1$)         & 0.124 & 0.688 & 0.133 & 0.688 & 0.129 & 0.688 \\
M9:  SIR (LDA) ($d=2$)         & 0.121 & 0.688 & 0.133 & 0.688 & 0.127 & 0.688 \\
M10: SAVE ($d=1$)        & 0.123 & 0.703 & 0.134 & 0.688 & 0.129 & 0.695 \\
M11: SAVE ($d=2$)
  & 0.123 & 0.688
  & 0.131 & 0.672
  & 0.127 & 0.680 \\
\bottomrule
\end{tabular}
\end{table}

The results in Table~\ref{tab:backtest_all} can be summarised in three groups.
The first group contains the non-SDR models, M1--M7. These models have combined RPS values between $0.209$ and $0.219$, with accuracies between $51\%$ and $55\%$. The differences within this group are small. Logistic regression, Poisson regression, ARIMA and NNAR variants, XGBoost, and the ensemble all perform similarly. The ensemble model, M7, is the best in this group, with a combined RPS of $0.209$. This suggests that, when the models use only the current Elo difference and recent form variables, the available predictive signal is limited.
The second group contains SDR models, M8--- M11. These models reduce the six-month Elo-difference history to one or two directions. This improves performance, lowering the combined RPS to $0.127$--$0.129$ and increasing accuracy to about $69\%$. The clearest pattern in Table~\ref{tab:backtest_all}, however, is the
gap between the two groups rather than the differences within them.
Every SDR model (M8--M11) has a lower combined RPS than every non-SDR
model (M1--M7). Most of this improvement is already realised with a single SDR
direction: the step from the current Elo difference to the reduced
Elo-history score (M3 to M8 or M10) accounts for nearly all of the gain,
while the second direction contributes the small additional refinement
described above. Because the SDR models share the goal model, the form
variables, and the neutral-venue indicator with the Elo-only Poisson
model (M3), the difference between them is due to the predictor alone:
M3 uses only the current Elo difference, whereas M8--M11 use a
low-dimensional summary of the six-month Elo-difference history. The
result indicates that the recent trajectory of the Elo difference
carries predictive information beyond its current value, and that
categorical SDR is an effective way to extract it.

\subsection{2026 World Cup Predictions}
\label{sec:pred2026}
Tournament forecasts are produced using M9 and M11, the best-performing models
in the backtests. We refit this model using the full training data from
January 2010 to the information barrier of June 1, 2026. This gives
$n=7{,}082$ matches involving World-Cup-qualified teams.
All tournament matches are treated as neutral-site matches. Since the
home and away labels are arbitrary in this setting, we symmetrise the
predicted probabilities for each pair of teams. That is, for each
unordered pair $\{h,a\}$,
\begin{equation*}
  P(h\text{ wins}) =
  \tfrac{1}{2}\bigl[ P(H\mid h\text{ home}) + P(A\mid a\text{ home}) \bigr].
\label{eq:sym}
\end{equation*}
Tournament outcomes are simulated using $N=5{,}000$ Monte Carlo replications. 
The group stage follows the official 48-team structure, that is, each group has four teams, each team plays the other three teams once, the top two teams in each group qualify automatically, and the eight best third-placed teams also advance to the Round of 32.
For matches in which both teams have sufficient Elo-history information, probabilities are obtained from \textbf{M9} and \textbf{M11}. For matches involving  Cura\c{c}ao,  which has insufficient information for stable SDR projection, probabilities are obtained from the Elo-only Poisson fallback model \textbf{M3}. This hybrid approach preserves the official tournament structure while avoiding unstable SDR scores for a team with limited historical information.

After the 32 qualifiers are determined in each simulation, we use the official FIFA Round-of-32 assignment table. 

In each simulated group-stage match, a scoreline is drawn from the fitted Poisson goal model. Teams are ranked first by points. Teams level on points are then separated by their head-to-head results among the tied teams (head-to-head points, then head-to-head goal difference, then head-to-head goals scored), and if a tie remains, by overall goal difference, overall goals scored, the fair-play score, and the FIFA World Ranking. The top two teams from each group advance to the Round of 32, and the eight best third-placed teams are selected across groups using points, goal difference, goals scored, and the remaining official criteria where available. The knockout stage then includes the Round of 32, Round of 16, quarter-finals, semi-finals, and final. If a knockout match is drawn after regular time in the simulation, the winner is selected by a 50/50 coin flip, representing extra time and penalties. 
 For each team $t$, we record the proportion of simulations in which it reaches each stage. This gives the estimated probabilities $\Pr(\mathrm{R32})$ through $\Pr(\mathrm{Champion})$. We also report each team’s projected group points, defined as its average number of group points across the simulations as 
\begin{align*}
   \widehat{\mathrm{xPts}}_{t}
= \frac{1}{N}
\sum_{s=1}^{N}
\left(3W_{t,s}+D_{t,s}\right),
\end{align*}
where $W_{t,s}$ and $D_{t,s}$ are team $t$'s wins and draws in group play in simulation $s$.  
Figure~\ref{fig:groups_SAVE} and Figure~\ref{fig:groups_sir} visualize the projected group standings under M11 and M9, respectively, and Table~\ref{tab:groupmatch_sir_save} reports all 72 group-stage matches under both models.  
The two reductions produce very similar group-stage forecasts. The projected group winner is the same in all twelve groups, and the two automatic qualifiers coincide in ten of them. The top two differ only in Group~J, where SAVE places Austria second and SIR places Algeria second, and in Group~K, where SAVE places Colombia second and SIR places DR~Congo second; in both cases the displaced team still advances as one of the eight best third-placed teams, so the qualifying field is almost unchanged. Across the full set of 32 teams reaching the Round of~32, the two models agree on 31 and differ in a single place: South~Korea qualifies as a best third-placed team under SAVE (M11), whereas DR~Congo takes that place under SIR (M9). At the match level, Table~\ref{tab:groupmatch_sir_save} shows that the 72 win/draw/loss probabilities are close, differing on average by about three to four percentage points in the home-win probability. The largest gaps occur in the most lopsided fixtures and the most draw-prone matches, where the second SAVE direction makes the probabilities slightly sharper; in evenly matched fixtures the two models almost coincide. No results from the 2026 FIFA World Cup final tournament are used in model fitting, feature construction, or simulation. 
\color{black}


\input{groups}

\begin{footnotesize}
\begin{longtable}{@{}lllcccccc@{}}
\caption{Group-stage match probabilities (\%) for the 2026 World Cup under the two categorical SDR models, SIR ($d{=}2$) and SAVE ($d{=}2$). $P(1)$, $P(\mathrm{X})$, and $P(2)$ are the probabilities of a Team~1 win, a draw, and a Team~2 win. Matches involving Cura\c{c}ao use the Elo-only Poisson fallback (M3).}
\label{tab:groupmatch_sir_save}\\
\toprule
\textbf{Grp} & \textbf{Team 1} & \textbf{Team 2}
  & \multicolumn{3}{c}{\textbf{SIR} ($d{=}2$)}
  & \multicolumn{3}{c}{\textbf{SAVE} ($d{=}2$)} \\
\cmidrule(lr){4-6}\cmidrule(lr){7-9}
 & & & $P(1)$ & $P(\mathrm{X})$ & $P(2)$ & $P(1)$ & $P(\mathrm{X})$ & $P(2)$ \\
\midrule
\endhead
\midrule
\multicolumn{9}{r}{\small\textit{Continued on next page.}}\\
\endfoot
\bottomrule
\endlastfoot
A & Mexico & South Korea & 63.8 & 22.5 & 13.7 & 60.9 & 23.6 & 15.5 \\
A & Mexico & South Africa & 64.1 & 22.3 & 13.6 & 66.8 & 21.1 & 12.2 \\
A & Mexico & Czechia & 50.8 & 26.9 & 22.3 & 51.8 & 26.6 & 21.5 \\
A & South Korea & South Africa & 36.0 & 28.2 & 35.8 & 41.4 & 28.2 & 30.4 \\
A & South Korea & Czechia & 24.3 & 26.7 & 49.0 & 27.6 & 27.8 & 44.7 \\
A & South Africa & Czechia & 24.2 & 26.6 & 49.2 & 22.8 & 26.5 & 50.7 \\
B & Canada & Bosnia & 41.0 & 29.0 & 30.0 & 46.3 & 28.1 & 25.6 \\
B & Canada & Qatar & 54.7 & 26.4 & 18.9 & 45.8 & 28.5 & 25.7 \\
B & Canada & Switzerland & 29.6 & 29.3 & 41.1 & 29.5 & 29.0 & 41.5 \\
B & Bosnia & Qatar & 50.1 & 26.7 & 23.2 & 35.4 & 28.7 & 35.9 \\
B & Bosnia & Switzerland & 25.0 & 26.9 & 48.0 & 20.8 & 25.8 & 53.4 \\
B & Qatar & Switzerland & 14.9 & 23.0 & 62.1 & 20.9 & 26.2 & 52.8 \\
C & Brazil & Morocco & 40.3 & 28.6 & 31.1 & 46.2 & 27.8 & 26.0 \\
C & Brazil & Haiti & 76.5 & 15.9 & 7.5 & 78.9 & 14.6 & 6.5 \\
C & Brazil & Scotland & 77.4 & 15.1 & 7.5 & 78.2 & 14.8 & 6.9 \\
C & Morocco & Haiti & 70.7 & 19.5 & 9.7 & 68.7 & 20.4 & 10.9 \\
C & Morocco & Scotland & 71.7 & 18.6 & 9.7 & 67.8 & 20.5 & 11.7 \\
C & Haiti & Scotland & 35.6 & 28.5 & 35.9 & 34.2 & 28.6 & 37.1 \\
D & United States & Paraguay & 15.3 & 21.9 & 62.8 & 15.6 & 22.6 & 61.9 \\
D & United States & Australia & 11.5 & 19.3 & 69.3 & 11.4 & 19.6 & 68.9 \\
D & United States & Turkey & 7.8 & 15.4 & 76.7 & 9.2 & 17.4 & 73.5 \\
D & Paraguay & Australia & 29.6 & 28.0 & 42.4 & 29.0 & 28.2 & 42.8 \\
D & Paraguay & Turkey & 23.2 & 26.3 & 50.5 & 25.2 & 27.2 & 47.6 \\
D & Australia & Turkey & 28.8 & 27.8 & 43.3 & 31.5 & 28.4 & 40.1 \\
E & Germany & Ivory Coast & 40.1 & 27.6 & 32.3 & 45.1 & 27.4 & 27.5 \\
E & Germany & Cura\c{c}ao & 71.3 & 18.4 & 10.3 & 71.3 & 18.4 & 10.3 \\
E & Germany & Ecuador & 45.2 & 27.9 & 26.9 & 42.6 & 28.4 & 29.0 \\
E & Ivory Coast & Cura\c{c}ao & 50.6 & 26.1 & 23.3 & 50.6 & 26.1 & 23.3 \\
E & Ivory Coast & Ecuador & 41.0 & 28.6 & 30.3 & 33.7 & 28.9 & 37.4 \\
E & Cura\c{c}ao & Ecuador & 10.7 & 19.0 & 70.3 & 10.7 & 19.0 & 70.3 \\
F & Netherlands & Japan & 27.4 & 27.8 & 44.7 & 28.4 & 28.2 & 43.4 \\
F & Netherlands & Sweden & 42.9 & 27.5 & 29.6 & 44.3 & 27.7 & 28.1 \\
F & Netherlands & Tunisia & 52.0 & 26.3 & 21.7 & 46.8 & 27.6 & 25.5 \\
F & Japan & Sweden & 52.1 & 25.9 & 22.0 & 52.3 & 26.0 & 21.6 \\
F & Japan & Tunisia & 61.1 & 23.5 & 15.5 & 54.8 & 25.7 & 19.5 \\
F & Sweden & Tunisia & 45.2 & 27.4 & 27.4 & 38.5 & 28.4 & 33.1 \\
G & Belgium & Egypt & 43.6 & 27.6 & 28.8 & 37.1 & 28.4 & 34.4 \\
G & Belgium & Iran & 52.4 & 25.9 & 21.7 & 50.5 & 26.6 & 22.9 \\
G & Belgium & New Zealand & 74.4 & 16.8 & 8.8 & 72.2 & 18.1 & 9.7 \\
G & Egypt & Iran & 43.6 & 28.7 & 27.8 & 48.5 & 27.6 & 23.9 \\
G & Egypt & New Zealand & 65.8 & 21.4 & 12.8 & 70.2 & 19.4 & 10.4 \\
G & Iran & New Zealand & 57.2 & 24.9 & 17.9 & 57.1 & 24.9 & 18.0 \\
H & Spain & Cape Verde & 86.2 & 10.1 & 3.7 & 88.4 & 8.7 & 2.9 \\
H & Spain & Saudi Arabia & 89.7 & 7.8 & 2.5 & 91.7 & 6.4 & 1.9 \\
H & Spain & Uruguay & 54.1 & 26.0 & 19.9 & 54.4 & 25.9 & 19.7 \\
H & Cape Verde & Saudi Arabia & 42.6 & 27.5 & 29.9 & 42.6 & 27.9 & 29.5 \\
H & Cape Verde & Uruguay & 10.2 & 18.8 & 71.0 & 8.5 & 17.2 & 74.4 \\
H & Saudi Arabia & Uruguay & 7.5 & 16.0 & 76.5 & 6.0 & 14.1 & 79.8 \\
I & France & Senegal & 61.7 & 22.8 & 15.5 & 62.5 & 22.7 & 14.9 \\
I & France & Iraq & 78.4 & 14.8 & 6.8 & 77.5 & 15.4 & 7.1 \\
I & France & Norway & 75.5 & 16.2 & 8.3 & 76.2 & 16.0 & 7.8 \\
I & Senegal & Iraq & 53.5 & 26.1 & 20.4 & 51.8 & 26.6 & 21.6 \\
I & Senegal & Norway & 49.6 & 26.7 & 23.7 & 49.9 & 26.8 & 23.3 \\
I & Iraq & Norway & 31.7 & 28.3 & 40.0 & 33.7 & 28.6 & 37.7 \\
J & Argentina & Algeria & 53.1 & 26.2 & 20.7 & 62.6 & 22.8 & 14.5 \\
J & Argentina & Austria & 54.2 & 25.4 & 20.5 & 57.9 & 24.4 & 17.7 \\
J & Argentina & Jordan & 71.4 & 18.6 & 10.0 & 87.0 & 9.6 & 3.4 \\
J & Algeria & Austria & 36.1 & 28.5 & 35.4 & 30.9 & 28.4 & 40.6 \\
J & Algeria & Jordan & 53.7 & 25.8 & 20.5 & 64.8 & 21.8 & 13.4 \\
J & Austria & Jordan & 54.1 & 25.2 & 20.7 & 70.2 & 19.1 & 10.7 \\
K & Portugal & DR Congo & 59.1 & 24.4 & 16.5 & 69.8 & 19.7 & 10.5 \\
K & Portugal & Uzbekistan & 64.9 & 21.8 & 13.4 & 67.1 & 20.8 & 12.1 \\
K & Portugal & Colombia & 61.0 & 23.1 & 15.9 & 59.6 & 23.8 & 16.6 \\
K & DR Congo & Uzbekistan & 40.2 & 29.5 & 30.3 & 32.3 & 29.4 & 38.3 \\
K & DR Congo & Colombia & 36.2 & 29.3 & 34.4 & 25.7 & 28.1 & 46.2 \\
K & Uzbekistan & Colombia & 31.5 & 28.7 & 39.8 & 28.5 & 28.3 & 43.2 \\
L & England & Croatia & 31.6 & 28.4 & 40.0 & 32.0 & 28.6 & 39.4 \\
L & England & Ghana & 67.5 & 20.5 & 12.0 & 69.6 & 19.6 & 10.8 \\
L & England & Panama & 39.1 & 28.8 & 32.1 & 41.5 & 28.6 & 29.9 \\
L & Croatia & Ghana & 72.3 & 17.9 & 9.8 & 73.6 & 17.4 & 9.0 \\
L & Croatia & Panama & 43.9 & 27.7 & 28.4 & 45.7 & 27.6 & 26.7 \\
L & Ghana & Panama & 14.1 & 21.8 & 64.2 & 14.1 & 22.0 & 63.9 \\
\end{longtable}
\end{footnotesize}

\begingroup \setlength{\tabcolsep}{2.2pt} 
\begin{footnotesize}
\begin{longtable}{@{}ll cccccc cccccc@{}}
\caption{2026 FIFA World Cup stage probabilities (\%), $N=5{,}000$ Monte Carlo
  simulations, for the two corrected categorical SDR models: SIR ($d=2$) and
  SAVE ($d=2$). R32 = reaching the Round of 32 (i.e.\ qualifying from the group
  stage); R16, QF, SF, Final, Champ as labelled. Cura\c{c}ao's matches use the
  Elo-only Poisson fallback (M3); all others use the stated SDR model. Teams
  ordered by SAVE champion probability.}
\label{tab:pred_2026_sir_save}\\
\toprule
& & \multicolumn{6}{c}{\textbf{SIR} ($d{=}2$)} & \multicolumn{6}{c}{\textbf{SAVE} ($d{=}2$)} \\
\cmidrule(lr){3-8}\cmidrule(lr){9-14}
\textbf{Team} & \textbf{Grp}
 & R32 & R16 & QF & SF & Fin & Chmp
 & R32 & R16 & QF & SF & Fin & Chmp \\
\midrule
\endfirsthead
\multicolumn{14}{l}{\small\textit{Table~\ref{tab:pred_2026_sir_save} continued.}}\\[2pt]
\toprule
& & \multicolumn{6}{c}{\textbf{SIR} ($d{=}2$)} & \multicolumn{6}{c}{\textbf{SAVE} ($d{=}2$)} \\
\cmidrule(lr){3-8}\cmidrule(lr){9-14}
\textbf{Team} & \textbf{Grp}
 & R32 & R16 & QF & SF & Fin & Chmp
 & R32 & R16 & QF & SF & Fin & Chmp \\
\midrule
\endhead
\midrule
\multicolumn{14}{r}{\small\textit{Continued on next page.}}\\
\endfoot
\bottomrule
\endlastfoot
Spain                  & H &  99.8 &  74.8 &  54.4 &  38.4 &  25.9 &  16.5 &  99.8 &  74.9 &  54.6 &  38.6 &  26.0 &  16.6 \\
Argentina              & J &  98.9 &  73.3 &  52.5 &  36.5 &  24.6 &  16.0 &  99.0 &  73.5 &  52.8 &  36.8 &  24.8 &  16.1 \\
France                 & I &  98.5 &  71.9 &  50.9 &  34.4 &  22.5 &  14.0 &  98.3 &  71.5 &  50.4 &  34.0 &  22.1 &  13.7 \\
Brazil                 & C &  98.2 &  64.4 &  39.9 &  23.6 &  13.3 &   6.8 &  98.0 &  64.2 &  39.8 &  23.7 &  13.4 &   6.9 \\
Portugal               & K &  95.9 &  61.5 &  37.0 &  21.2 &  11.3 &   5.6 &  95.2 &  59.7 &  35.2 &  19.7 &  10.3 &   5.1 \\
England                & L &  97.0 &  59.3 &  34.2 &  18.2 &   9.3 &   4.6 &  97.2 &  60.6 &  35.6 &  19.3 &  10.0 &   5.0 \\
Netherlands            & F &  94.2 &  57.7 &  32.9 &  17.9 &   9.0 &   4.5 &  94.6 &  57.8 &  33.1 &  17.9 &   9.1 &   4.5 \\
Germany                & E &  92.5 &  56.1 &  31.2 &  16.5 &   8.1 &   3.4 &  92.7 &  56.1 &  31.3 &  16.6 &   8.1 &   3.5 \\
Japan                  & F &  90.5 &  53.5 &  29.8 &  15.6 &   7.3 &   3.3 &  90.5 &  53.2 &  29.5 &  15.3 &   7.1 &   3.2 \\
Colombia               & K &  74.8 &  43.5 &  23.6 &  12.4 &   6.2 &   2.7 &  76.6 &  45.5 &  25.5 &  13.8 &   6.9 &   3.2 \\
Croatia                & L &  92.0 &  51.8 &  26.9 &  12.6 &   5.7 &   2.4 &  91.9 &  51.9 &  27.2 &  12.9 &   5.8 &   2.4 \\
Morocco                & C &  91.7 &  51.2 &  26.9 &  13.0 &   5.8 &   2.5 &  91.0 &  50.6 &  26.4 &  12.8 &   5.5 &   2.3 \\
Ecuador                & E &  78.7 &  43.0 &  22.2 &  10.7 &   4.9 &   2.0 &  79.2 &  43.4 &  22.5 &  10.9 &   5.0 &   2.0 \\
Belgium                & G &  93.5 &  51.2 &  25.9 &  12.2 &   5.1 &   2.1 &  93.3 &  50.4 &  25.1 &  11.7 &   4.8 &   1.9 \\
Uruguay                & H &  77.5 &  40.5 &  19.9 &   8.8 &   3.9 &   1.5 &  75.9 &  39.2 &  19.0 &   8.3 &   3.6 &   1.4 \\
Senegal                & I &  80.2 &  41.8 &  20.0 &   9.0 &   3.8 &   1.5 &  79.1 &  40.7 &  19.5 &   8.6 &   3.6 &   1.4 \\
Mexico                 & A &  94.9 &  46.5 &  21.3 &   9.1 &   3.6 &   1.3 &  94.3 &  45.6 &  20.6 &   8.7 &   3.4 &   1.2 \\
Switzerland            & B &  82.6 &  40.3 &  18.4 &   7.5 &   2.9 &   1.1 &  83.0 &  41.0 &  19.0 &   7.7 &   3.1 &   1.1 \\
Turkey                 & D &  67.6 &  34.4 &  16.2 &   7.1 &   2.8 &   1.0 &  66.6 &  34.2 &  16.2 &   7.1 &   2.8 &   1.0 \\
Austria                & J &  70.3 &  35.3 &  15.9 &   6.5 &   2.6 &   1.0 &  69.6 &  34.4 &  15.3 &   6.0 &   2.4 &   0.9 \\
Algeria                & J &  75.1 &  36.2 &  16.2 &   6.6 &   2.4 &   0.8 &  76.1 &  37.0 &  16.6 &   6.7 &   2.4 &   0.8 \\
Iran                   & G &  76.5 &  36.7 &  16.5 &   6.5 &   2.3 &   0.7 &  77.0 &  36.8 &  16.4 &   6.6 &   2.2 &   0.8 \\
Norway                 & I &  51.7 &  25.4 &  11.2 &   4.7 &   1.6 &   0.6 &  54.2 &  27.4 &  12.6 &   5.5 &   2.0 &   0.7 \\
Canada                 & B &  93.9 &  41.0 &  16.5 &   6.1 &   1.9 &   0.5 &  94.2 &  41.8 &  17.0 &   6.4 &   2.0 &   0.6 \\
Australia              & D &  68.0 &  32.4 &  13.9 &   5.6 &   2.0 &   0.7 &  65.6 &  30.6 &  12.8 &   5.1 &   1.9 &   0.6 \\
Paraguay               & D &  71.0 &  30.8 &  12.4 &   4.6 &   1.6 &   0.5 &  70.5 &  30.6 &  12.4 &   4.6 &   1.5 &   0.5 \\
Egypt                  & G &  76.8 &  33.1 &  13.0 &   4.7 &   1.5 &   0.5 &  75.2 &  31.2 &  11.9 &   4.2 &   1.2 &   0.4 \\
South Korea            & A &  81.6 &  32.2 &  11.6 &   3.6 &   1.1 &   0.3 &  82.8 &  33.6 &  12.5 &   4.1 &   1.2 &   0.4 \\
Ivory Coast            & E &  64.9 &  27.6 &  10.9 &   4.0 &   1.4 &   0.4 &  64.0 &  26.7 &  10.2 &   3.7 &   1.2 &   0.3 \\
Panama                 & L &  52.5 &  22.6 &   8.8 &   3.0 &   1.0 &   0.3 &  52.8 &  22.9 &   9.0 &   3.2 &   1.1 &   0.3 \\
Uzbekistan             & K &  47.4 &  18.7 &   6.9 &   2.1 &   0.7 &   0.2 &  48.0 &  19.2 &   7.2 &   2.2 &   0.8 &   0.2 \\
United States          & D &  66.1 &  24.2 &   8.1 &   2.6 &   0.6 &   0.1 &  69.7 &  26.6 &   9.3 &   3.1 &   0.9 &   0.2 \\
Tunisia                & F &  34.2 &  13.9 &   5.3 &   1.8 &   0.5 &   0.1 &  35.2 &  14.2 &   5.3 &   1.8 &   0.6 &   0.1 \\
Czechia                & A &  52.1 &  18.9 &   6.1 &   1.9 &   0.4 &   0.1 &  51.4 &  18.6 &   6.0 &   1.8 &   0.4 &   0.1 \\
Sweden                 & F &  46.7 &  18.1 &   6.4 &   1.9 &   0.6 &   0.1 &  45.0 &  16.9 &   5.7 &   1.6 &   0.5 &   0.1 \\
DR Congo               & K &  47.9 &  17.1 &   5.3 &   1.7 &   0.5 &   0.1 &  46.7 &  16.6 &   5.1 &   1.6 &   0.5 &   0.1 \\
Iraq                   & I &  33.5 &  11.5 &   3.7 &   1.1 &   0.2 &   0.1 &  32.6 &  11.1 &   3.6 &   1.1 &   0.2 &   0.1 \\
Scotland               & C &  31.7 &  10.9 &   3.2 &   0.9 &   0.3 &   0.1 &  32.1 &  11.3 &   3.5 &   1.0 &   0.3 &   0.1 \\
Haiti                  & C &  41.1 &  13.0 &   3.7 &   0.9 &   0.2 &   0.0 &  41.8 &  13.6 &   3.8 &   0.9 &   0.2 &   0.0 \\
Cura\c{c}ao            & E &  36.2 &  11.0 &   3.1 &   0.8 &   0.2 &   0.0 &  36.2 &  11.0 &   3.1 &   0.7 &   0.2 &   0.0 \\
South Africa           & A &  38.2 &   9.5 &   2.2 &   0.4 &   0.1 &   0.0 &  38.7 &   9.9 &   2.3 &   0.4 &   0.1 &   0.0 \\
New Zealand            & G &  24.7 &   7.8 &   2.3 &   0.6 &   0.1 &   0.0 &  25.8 &   8.1 &   2.3 &   0.6 &   0.1 &   0.0 \\
Jordan                 & J &  21.6 &   7.2 &   2.1 &   0.5 &   0.1 &   0.0 &  21.1 &   6.9 &   2.0 &   0.5 &   0.1 &   0.0 \\
Bosnia and Herzegovina & B &  58.2 &  14.3 &   3.1 &   0.6 &   0.1 &   0.0 &  58.8 &  14.7 &   3.3 &   0.6 &   0.1 &   0.0 \\
Cape Verde             & H &  48.4 &  12.2 &   2.7 &   0.5 &   0.1 &   0.0 &  48.9 &  12.6 &   2.9 &   0.6 &   0.1 &   0.0 \\
Qatar                  & B &  33.5 &   7.4 &   1.4 &   0.2 &   0.0 &   0.0 &  32.3 &   7.1 &   1.3 &   0.2 &   0.0 &   0.0 \\
Saudi Arabia           & H &  33.1 &   8.5 &   2.1 &   0.5 &   0.0 &   0.0 &  33.8 &   8.8 &   2.2 &   0.4 &   0.1 &   0.0 \\
Ghana                  & L &  24.2 &   5.9 &   1.2 &   0.2 &   0.0 &   0.0 &  23.6 &   5.8 &   1.1 &   0.2 &   0.0 &   0.0 \\
\end{longtable}
\end{footnotesize}

Table~\ref{tab:pred_2026_sir_save} reports each team's probability of reaching every knockout round under the two models, obtained from a Monte Carlo simulation of the full tournament ($N=5,000$). Unlike Table~\ref{tab:groupmatch_sir_save}, which gives the win, draw, and loss probabilities of individual fixtures, and Figures~\ref{fig:groups_SAVE} and~\ref{fig:groups_sir}, which rank teams within their groups, this table is organised by team alone: each entry is a team's overall chance of advancing to a given round, aggregated over all the matches and paths it could take, rather than the outcome of any single game or its standing within a group. Both models make Spain the clear favourite, Argentina and France are effectively level for second place. For the strong teams, the two models track each other closely at every stage, and the top of the championship ranking is the same. The visible differences between SIR and SAVE are confined to the weaker and borderline teams.

Figures~\ref{fig:bracket} (SAVE) and~\ref{fig:bracket_sir} (SIR) show illustrative knockout brackets under the SAVE and SIR models, with the corresponding tie-by-tie win probabilities in Table~\ref{tab:knockout_sir_save}. Once the 32 qualifiers are determined,
teams are placed on the official FIFA Round-of-32 slot template: group winners, runners-up and the eight best third-placed teams occupy fixed positions, so that two group winners never meet in the Round of 32 and no team faces a side from its own group before the final. Each tie is then resolved by the model's win probability, yielding one illustrative path to the title under each model. These brackets follow the official slot structure but remain model projections conditional on the predicted group outcomes, not a claim about the realized draw or result. Because the official template places France (winner~I) and Spain
(winner~H) in the same half, the two can meet only in the semifinals rather
than the final; the higher-rated of the two (Spain) advances along this path
under both models. 

  Table~\ref{tab:pred_2026_sir_save} gives the primary tournament forecast: it reports marginal probabilities averaged over all $N=5,000$ Monte Carlo simulations. The illustrative brackets answer a different question from the marginal simulation probabilities and therefore need not agree exactly on finalists or stage probabilities. On the official-template, both SDR models give the same outcome---champion \textbf{Spain}, reached through a Spain--Argentina final. This agrees with Table~\ref{tab:pred_2026_sir_save}, where Spain has the highest championship probability.  France also has a high marginal probability of reaching the final, but in the representative path, Spain and France are placed in the same half and meet in the semi-final; Spain advances, while Argentina reaches the final from the opposite half.  Thus, Table~\ref{tab:pred_2026_sir_save} should be used for the overall team ranking, while Figures~\ref{fig:bracket}  and~\ref{fig:bracket_sir} and Table~\ref{tab:knockout_sir_save} summarize one representative knockout path under the seeded-bracket approximation.

\input{bracket.tex}

\begingroup \setlength{\tabcolsep}{2.2pt}
\begin{footnotesize}
\begin{longtable}{@{}llcc@{}}
\caption{Projected knockout ties for the two categorical SDR models, SIR ($d=2$) and SAVE ($d=2$), with teams placed on the official FIFA Round-of-32 slot template. Win\% is the probability that the first-named team advances, including extra time and penalties. The ties are conditional on the predicted group outcomes and are not a claim about the realized draw or result. Rows marked $^{\dagger}$ occur under only one model, where the two advance different teams upstream.}
\label{tab:knockout_sir_save}\\
\toprule
\textbf{Round} & \textbf{Match} & \textbf{SIR Win\%} & \textbf{SAVE Win\%} \\
\midrule
\endfirsthead
\multicolumn{4}{l}{\small\textit{Table~\ref{tab:knockout_sir_save} continued.}}\\[2pt]
\toprule
\textbf{Round} & \textbf{Match} & \textbf{SIR Win\%} & \textbf{SAVE Win\%} \\
\midrule
\endhead
\midrule
\multicolumn{4}{r}{\small\textit{Continued on next page.}}\\
\endfoot
\bottomrule
\multicolumn{4}{l}{\small$^{\dagger}$Tie that occurs under only one model (the models advance different teams upstream).}
\endlastfoot
Round of 32 & \textbf{Germany} vs South Korea$^{\dagger}$ & -- & \textbf{68.6} / 31.4 \\
 & \textbf{France} vs Paraguay & \textbf{78.9} / 21.1 & \textbf{78.6} / 21.4 \\
 & Czechia vs \textbf{Canada} & 42.6 / \textbf{57.4} & 42.0 / \textbf{58.0} \\
 & \textbf{Japan} vs Morocco & \textbf{53.7} / 46.3 & \textbf{53.6} / 46.4 \\
 & Colombia vs \textbf{England}$^{\dagger}$ & -- & 48.0 / \textbf{52.0} \\
 & \textbf{Spain} vs Austria$^{\dagger}$ & -- & \textbf{77.0} / 23.0 \\
 & \textbf{Turkey} vs Bosnia$^{\dagger}$ & -- & \textbf{77.1} / 22.9 \\
 & \textbf{Belgium} vs Ivory Coast$^{\dagger}$ & -- & \textbf{61.9} / 38.1 \\
 & \textbf{Brazil} vs Netherlands & \textbf{54.6} / 45.4 & \textbf{54.8} / 45.2 \\
 & \textbf{Ecuador} vs Senegal & \textbf{52.9} / 47.1 & \textbf{53.6} / 46.4 \\
 & \textbf{Mexico} vs Sweden & \textbf{61.3} / 38.7 & \textbf{61.9} / 38.1 \\
 & \textbf{Croatia} vs Algeria$^{\dagger}$ & -- & \textbf{57.3} / 42.7 \\
 & \textbf{Argentina} vs Uruguay & \textbf{72.4} / 27.6 & \textbf{73.1} / 26.9 \\
 & \textbf{Australia} vs Egypt & \textbf{54.8} / 45.2 & \textbf{55.6} / 44.4 \\
 & \textbf{Switzerland} vs Iran & \textbf{51.2} / 48.8 & \textbf{51.9} / 48.1 \\
 & \textbf{Portugal} vs Panama & \textbf{71.4} / 28.6 & \textbf{69.8} / 30.2 \\
 & \textbf{Germany} vs Bosnia$^{\dagger}$ & \textbf{83.7} / 16.3 & -- \\
 & DR Congo vs \textbf{England}$^{\dagger}$ & 24.5 / \textbf{75.5} & -- \\
 & \textbf{Spain} vs Algeria$^{\dagger}$ & \textbf{77.0} / 23.0 & -- \\
 & \textbf{Turkey} vs Ivory Coast$^{\dagger}$ & \textbf{59.2} / 40.8 & -- \\
 & \textbf{Belgium} vs Austria$^{\dagger}$ & \textbf{54.7} / 45.3 & -- \\
 & Croatia vs \textbf{Colombia}$^{\dagger}$ & 46.7 / \textbf{53.3} & -- \\
\midrule
Round of 16 & Germany vs \textbf{France} & 34.7 / \textbf{65.3} & 35.1 / \textbf{64.9} \\
 & Canada vs \textbf{Japan} & 34.6 / \textbf{65.4} & 35.8 / \textbf{64.2} \\
 & England vs \textbf{Spain} & 34.3 / \textbf{65.7} & 35.3 / \textbf{64.7} \\
 & Turkey vs \textbf{Belgium} & 47.4 / \textbf{52.6} & 48.3 / \textbf{51.7} \\
 & \textbf{Brazil} vs Ecuador & \textbf{60.3} / 39.7 & \textbf{60.1} / 39.9 \\
 & Mexico vs \textbf{Croatia}$^{\dagger}$ & -- & 42.8 / \textbf{57.2} \\
 & \textbf{Argentina} vs Australia & \textbf{77.0} / 23.0 & \textbf{77.9} / 22.1 \\
 & Switzerland vs \textbf{Portugal} & 34.2 / \textbf{65.8} & 35.9 / \textbf{64.1} \\
 & Mexico vs \textbf{Colombia}$^{\dagger}$ & 40.5 / \textbf{59.5} & -- \\
\midrule
Quarter-finals & \textbf{France} vs Japan & \textbf{65.0} / 35.0 & \textbf{65.1} / 34.9 \\
 & \textbf{Spain} vs Belgium & \textbf{72.7} / 27.3 & \textbf{73.6} / 26.4 \\
 & \textbf{Brazil} vs Croatia$^{\dagger}$ & -- & \textbf{60.0} / 40.0 \\
 & \textbf{Argentina} vs Portugal & \textbf{62.1} / 37.9 & \textbf{63.4} / 36.6 \\
 & \textbf{Brazil} vs Colombia$^{\dagger}$ & \textbf{56.8} / 43.2 & -- \\
\midrule
Semi-finals & France vs \textbf{Spain} & 47.2 / \textbf{52.8} & 46.7 / \textbf{53.3} \\
 & Brazil vs \textbf{Argentina} & 39.1 / \textbf{60.9} & 39.1 / \textbf{60.9} \\
\midrule
Third place & \textbf{France} vs Brazil & \textbf{58.9} / 41.1 & \textbf{58.4} / 41.6 \\
\midrule
Final & \textbf{Spain} vs Argentina & \textbf{50.7} / 49.3 & \textbf{50.7} / 49.3 \\
\end{longtable}
\end{footnotesize}
\endgroup

\section{Conclusion}
\label{sec:conclusion}

We developed a forecasting framework for the 2026 FIFA World Cup based on recent Elo-rating histories. Instead of using only the current Elo difference, we constructed a six-month lagged Elo-difference vector and reduced it using categorical sufficient dimension reduction. The resulting SIR and SAVE scores were then used in a Poisson double-regression model for home and away goals.

In out-of-sample backtests on the 2018 and 2022 World Cups, the SDR-based Poisson models performed better than the non-SDR baselines. The strongest models were SIR with two directions (M9) and SAVE with two directions (M11), both with combined RPS about 0.127 and accuracy around 68\%. This suggests that recent Elo-history information contains a predictive signal beyond the current Elo difference alone. The main improvement comes from replacing the single current Elo difference with a low-dimensional summary of the recent Elo trajectory; the second direction provides a smaller additional refinement.

For the 2026 tournament forecast, we retained all 48 teams. Matches involving Cura\c{c}ao were modeled using the Elo-only Poisson fallback because it has insufficient history for a stable SDR projection. Both SDR models give similar tournament forecasts and identify Spain as the favourite, followed by Argentina, France, Brazil, and Portugal. Future work could improve the framework by incorporating bookmaker odds, player-level information, squad updates, or dynamic team-strength models that evolve jointly across teams.

\bibliographystyle{apalike}
\bibliography{Ref1}

\end{document}